\documentclass[twocolumn,aps,showpacs,floatfix,prc]{revtex4}
\usepackage[dvips]{epsfig}
\begin{document}

\title{ Stability of the $\beta$-equilibrated dense matter and core-crust transition in neutron stars }
\author{ Debasis Atta$^1$ and D.N. Basu$^2$ }

\affiliation{ Variable  Energy  Cyclotron  Centre, 1/AF Bidhan Nagar, Kolkata 700 064, India }
\email[E-mail 1: ]{datta@vecc.gov.in}
\email[E-mail 2: ]{dnb@vecc.gov.in}

\date{\today }

\begin{abstract}

    The stability of the $\beta$-equilibrated dense nuclear matter is analyzed with respect to the thermodynamic stability conditions. Based on the density dependent M3Y effective nucleon-nucleon interaction, the effects of the nuclear incompressibility on the proton fraction in neutron stars and the location of the inner edge of their crusts and core-crust transition density and pressure are investigated. The high-density behavior of symmetric and asymmetric nuclear matter satisfies the constraints from the observed flow data of heavy-ion collisions. The neutron star properties studied using $\beta$-equilibrated neutron star matter obtained from this effective interaction for a pure hadronic model agree with the recent observations of the massive compact stars. The density, pressure and proton fraction at the inner edge separating the liquid core from the solid crust of neutron stars are determined to be $\rho_t=$ 0.0938 fm$^{-3}$, P$_t=$ 0.5006 MeV fm$^{-3}$ and x$_{p(t)}=$ 0.0308, respectively.

\vskip 0.2cm
\noindent
{\it Keywords}: Nuclear EoS; Symmetry energy; Neutron Star; Core-crust transition.  

\end{abstract}

\pacs{ 21.65.-f, 26.60.-c, 26.60.Dd, 26.60.Gj, 97.60.Jd, 21.30.Fe }   
\maketitle

\noindent
\section{Introduction}
\label{section1}

    The equation of state (EoS) of nuclear matter under exotic conditions is an indispensable tool for the understanding of the nuclear force and for astrophysical applications. This implies knowledge of EoS at high isospin asymmetries and for a wide density range (both for subsaturation and suprasaturation densities). In order to ascertain our knowledge on the nature of matter under extreme conditions, neutron stars are among the most mysterious objects in the universe that provide natural laboratory. Understanding their structures and properties has long been a very challenging task for both the astrophysics and the nuclear physics community \cite{La04}. 
   
    One of the most important predictions of an EoS is the location of the inner edge of a neutron star crust. Knowledge of the properties of the crust plays an important role in understanding many astrophysical observations \cite{Ba71,Ba71a,Pe95,Pe95a,La01,La07,St05,Li99,Ho04,Ho04a,Bu06,Ow05}. The inner crust spans the region from the neutron drip point to the inner edge separating the solid crust from the homogeneous liquid core. While the neutron drip density $\rho_d$ is relatively well determined to be about 4.3$\times$10$^{11}$ g cm$^{-3}$ \cite{Ru06}, the transition density $\rho_t$ at the inner edge is still largely uncertain mainly because of limited knowledge on EoS, especially the density dependence of the symmetry energy, of neutron-rich nuclear matter \cite{La01,La07}. At the inner edge a phase transition occurs from the high-density homogeneous matter to the inhomogeneous one at lower densities. The transition density takes its critical value $\rho_t$ when the uniform neutron-proton-electron matter (npe) becomes unstable with respect to the separation into two coexisting phases (one corresponding to nuclei, the other to a nucleonic sea) \cite{La07}.

    In general, the determination of the transition density $\rho_t$ itself is a very complicated problem because the inner crust may have a very complicated structure. A well established approach is to find the density at which the uniform liquid first becomes unstable against small-amplitude density fluctuations, indicating the formation of nuclear clusters. This approach includes the dynamical method \cite{Ba71,Ba71a,Pe95,Pe95a,Do00,Oy07,Du07,Xu09,Xu09a}, the thermodynamical one \cite{La07,Ku04,Ku07,Wo08} and the random phase approximation (RPA) \cite{Ho01,Ca03}.
    
    In the present work, using the EoS for neutron-rich nuclear matter constrained by the recent isospin diffusion data from heavy-ion reactions in the same subsaturation density range as the neutron star crust, the inner edge of neutron star crusts is determined. For the EoS used in the present work, which is obtained from the density dependent M3Y effective nucleon-nucleon interaction (DDM3Y), the incompressibility $K_\infty$ for the symmetric nuclear matter (SNM), nuclear symmetry energy $E_{sym}(\rho_0)$ at saturation density $\rho_0$, the isospin dependent part $K_\tau$ of the isobaric incompressibility and the slope $L$ are all in excellent agreement with the constraints recently extracted from measured isotopic dependence of the giant monopole resonances in even-A Sn isotopes, from the neutron skin thickness of nuclei, and from analyses of experimental data on isospin diffusion and isotopic scaling in intermediate energy heavy-ion collisions \cite{Ch09,Ba09}. The core-crust transition in neutron stars is determined by analyzing the stability of the $\beta$-equilibrated dense nuclear matter with respect to the thermodynamic stability conditions. 

\noindent
\section{Intrinsic stability of the neutron star matter under $\beta$-equilibrium}
\label{section2}

    Inner edge of the neutron star crusts corresponds to a phase transition from the homogeneous matter at high densities to the inhomogeneous matter at low densities. In principle, the inner edge can be located by a detailed comparison of the relevant properties of the nonuniform solid crust and the uniform liquid core  consisting mainly of the npe matter. However, this procedure is impracticable as the inner crust may contain nuclei having very complicated geometries, usually known as the "nuclear pasta" \cite{La04,Ho04,Ho04a,Ra83,Oy93,St08}. Moreover, the core-crust transition is expected to be a very weak first-order phase transition and model calculations lead to very small density discontinuities at the transition \cite{Pe95,Do00,Do01,Ca03}. In practice, therefore, a good approximation is to search for the density at which the uniform liquid first becomes unstable against small amplitude density fluctuations with clusterization. This approximation has been shown to produce a very small error for the actual core-crust transition density and would yield the exact transition density for a second-order phase transition \cite{Pe95,Do00,Do01,Ca03}. Here, we use the thermodynamical method for analyzing the stability of the neutron star matter under $\beta$-equilibrium.

\subsection{The equation of state}

    The nuclear matter EoS is calculated using the isoscalar and the isovector \cite{La62,Sa83} components of M3Y interaction along with density dependence. The density dependence of this DDM3Y effective interaction is completely determined from nuclear matter calculations. The equilibrium density of the nuclear matter is determined by minimizing the energy per nucleon. The energy variation of the zero range potential is treated accurately by allowing it to vary freely with the kinetic energy part $\epsilon^{kin}$ of the energy per nucleon $\epsilon$ over the entire range of $\epsilon$. This is not only more plausible, but also yields excellent result for the incompressibility $K_\infty$ of the SNM which does not suffer from the superluminosity problem. 
        
    In a Fermi gas model of interacting neutrons and protons, with isospin asymmetry $X= \frac{\rho_n - \rho_p} {\rho_n + \rho_p},~~~~\rho = \rho_n+\rho_p,$ where $\rho_n$, $\rho_p$ and $\rho$ are the neutron, proton and nucleonic densities respectively, the energy per nucleon for isospin asymmetric nuclear matter can be derived as \cite{BCS08}

\begin{equation}
 \epsilon(\rho,X) = [\frac{3\hbar^2k_F^2}{10m}] F(X) + (\frac{\rho J_v C}{2}) (1 - \beta\rho^n)  
\label{seqn1}
\end{equation}
\noindent
where $m$ is the nucleonic mass, $k_F$=$(1.5\pi^2\rho)^{\frac{1}{3}}$ which equals Fermi momentum in case of SNM, the kinetic energy per nucleon $\epsilon^{kin}$=$[\frac{3\hbar^2k_F^2}{10m}] F(X)$ with $F(X)$=$[\frac{(1+X)^{5/3} + (1-X)^{5/3}}{2}]$ and $J_v$=$J_{v00} + X^2 J_{v01}$, $J_{v00}$ and $J_{v01}$ represent the volume integrals of the isoscalar and the isovector parts of the M3Y interaction. The isoscalar $t_{00}^{M3Y}$ and the isovector $t_{01}^{M3Y}$ components of M3Y interaction potential are given by $t_{00}^{M3Y}(s, \epsilon)$=7999$\frac{\exp( - 4s)}{4s}$-$2134\frac{\exp( - 2.5s)}{2.5s}$+$J_{00}$(1-$\alpha\epsilon$)$\delta(s)$, $t_{01}^{M3Y}(s, \epsilon)$=-4886$\frac{\exp( - 4s)}{4s}$+$1176\frac{\exp( - 2.5s)}{2.5s}$+$J_{01}$(1-$\alpha\epsilon$)$\delta(s)$ $J_{00}$=-276 MeVfm$^3$, $J_{01}$=228 MeVfm$^3$, $\alpha=0.005$MeV$^{-1}$. The DDM3Y effective NN interaction is given by $v_{0i}(s,\rho, \epsilon) = t_{0i}^{M3Y}(s, \epsilon) g(\rho)$ where the density dependence $g(\rho) = C (1 - \beta \rho^n)$ with $C$ and $\beta$ being the constants of density dependence.

    The Eq.(1) can be differentiated with respect to $\rho$ to yield equation for $X=0$:  
    
\begin{equation}
 \frac{\partial\epsilon}{\partial\rho} = [\frac{\hbar^2k_F^2}{5m\rho}] + \frac{J_{v00} C}{2} [1 - (n+1)\beta\rho^n] 
-\alpha J_{00} C [1 - \beta\rho^n]  [\frac{\hbar^2k_F^2}{10m}].
\label{seqn2}
\end{equation}
\noindent
The equilibrium density of the cold SNM is determined from the saturation condition. Then Eq.(1) and Eq.(2) with the saturation condition $\frac{\partial\epsilon}{\partial\rho} = 0$ at $\rho = \rho_{0}$, $\epsilon = \epsilon_{0}$ can be solved simultaneously for fixed values of the saturation energy per nucleon $\epsilon_0$ and the saturation density $\rho_{0}$ of the cold SNM to obtain the values of $\beta$ and $C$. The constants of density dependence $\beta$ and $C$, thus obtained, are given by 

\begin{equation}
 \beta = \frac{[(1-p)+(q-\frac{3q}{p})]\rho_{0}^{-n}}{[(3n+1)-(n+1)p+(q-\frac{3q}{p})]}
\label{seqn3}
\end{equation} 
\noindent
where $p$=$\frac{[10m\epsilon_0]}{[\hbar^2k_{F_0}^2]}$, $q$=$\frac{2\alpha\epsilon_0J_{00}}{J^0_{v00}}$, $J^0_{v00}$=$J_{v00}(\epsilon^{kin}_0)$ implying $J_{v00}$ at $\epsilon^{kin}$=$\epsilon^{kin}_0$, the kinetic energy part of the saturation energy per nucleon of SNM,  $k_{F_0}$=$[1.5\pi^2\rho_0]^{1/3}$ and 

\begin{equation}
 C = -\frac{[2\hbar^2k_{F_0}^2] }{ 5mJ^0_{v00} \rho_0 [1 - (n+1)\beta\rho_0^n -\frac{q\hbar^2k_{F_0}^2 (1-\beta\rho_0^n)}{10m\epsilon_0}]},
\label{seqn4}
\end{equation} 
\noindent
respectively. It is quite obvious that the constants of density dependence $C$ and $\beta$ obtained by this method depend on the saturation energy per nucleon $\epsilon_0$, the saturation density $\rho_{0}$, the index $n$ of the density dependent part and on the strengths of the M3Y interaction through the volume integral $J^0_{v00}$. 

    The calculations are performed using the values of the saturation density $\rho_0$=0.1533 fm$^{-3}$ \cite{Sa89} and the saturation energy per nucleon $\epsilon_0$=-15.26 MeV \cite{CB06} for the SNM obtained from the co-efficient of the volume term of Bethe-Weizs\"acker mass formula which is evaluated by fitting the recent experimental and estimated atomic mass excesses from Audi-Wapstra-Thibault atomic mass table \cite{Au03} by minimizing the mean square deviation incorporating correction for the electronic binding energy \cite{Lu03}. In a similar recent work, including surface symmetry energy term, Wigner term, shell correction and proton form factor correction to Coulomb energy also, $a_v$ turns out to be 15.4496 MeV and 14.8497 MeV when $A^0$ and $A^{1/3}$ terms are also included \cite{Ro06}. Using the usual values of $\alpha$=0.005 MeV$^{-1}$ for the parameter of energy dependence of the zero range potential and $n$=2/3, the values obtained for the constants of density dependence $C$ and $\beta$ and the SNM incompressibility $K_\infty$ are 2.2497, 1.5934 fm$^2$ and 274.7 MeV, respectively. The saturation energy per nucleon is the volume energy coefficient and the value of -15.26$\pm$0.52 MeV covers, more or less, the entire range of values obtained for $a_v$ for which now the values of $C$=2.2497$\pm$0.0420, $\beta$=1.5934$\pm$0.0085 fm$^2$ and the SNM incompressibility $K_\infty$=274.7$\pm$7.4 MeV.  

\subsection{ Intrinsic stability of single phase under $\beta$-equilibrium and the core-crust transition }

    The basic equation in neutron star matter research is the shape of the relationship between the pressure and energy density $P=P(\varepsilon)$, usually called the equation of state. At the zero temperature, the state of neutron star matter should be uniquely described by the quantities that are conserved by the process leading to equilibrium. Stable high density nuclear matter must be in chemical equilibrium for all types of reactions including the weak interactions, while the beta decay and orbital electron capture takes place simultaneously. For the $\beta$-equilibrated neutron star matter we have free neutron decay $n\rightarrow p+\beta^- +\overline{\nu_e}$ which are governed by weak interaction and the electron capture process $p+\beta^-\rightarrow n+\nu_e$. Both types of reactions change the electron fraction and thus affect the EoS. Here we assume that neutrinos generated in these reactions leave the system. The absence of neutrino has a dramatic effect on the equation of state and mainly induces a significant change on the values of proton fraction $x_p$. The absence of neutrino implies that
    
\begin{eqnarray}
\mu= {\mu_n}-{\mu_p}=\mu_e
\label{seqn5}
\end{eqnarray}
where $\mu_e$, $\mu_n$ and $\mu_p$ are the chemical potentials for electron, neutron and proton, respectively. 

    The baryon number $B$ is conserved by this type of reaction so the energy density $\varepsilon$ and pressure $P$ should be function of baryon number density $\rho$. We assume that the matter is electrically neutral and spatially homogeneous. The star as a whole is electrically neutral but the matter does not need to be locally neutral. So the thermodynamic state of a given phase is described by two quantities: baryon number B and charge Q where Q is the sum of all charges. The total energy U then becomes a function of U(V,B,Q). To consider stability of a single phase one need to introduce local quantities $\epsilon=\frac{U}{B}$. The energy per particle $\epsilon$ then becomes a function of other local quantities taken per baryon number $v=\frac{V}{B} $ and $x_p=\frac{Q}{B}$. The first principle of thermodynamics takes the following form: 

\begin{equation}
d\epsilon=-Pdv-\mu dx_p
\label{seqn6}
\end{equation}
where $P$ is the pressure and $\mu$ is the chemical potential of an electric charge. The stability of any single phase, also called intrinsic stability, is ensured by convexity of $\epsilon(v,x_p)$. The thermodynamical inequalities allows us to express the requirement in terms of following inequalities:

\begin{equation}
-(\frac{\partial P}{\partial v})_{x_p}>0,  ~~~~  -(\frac{\partial \mu}{\partial x_p})_P>0
\label{seqn7}
\end{equation}
One may find another pair of inequalities that are equivalent to above equations:

\begin{equation}
-(\frac{\partial P}{\partial v})_\mu>0,  ~~~~  -(\frac{\partial \mu}{\partial x_p})_v>0
\label{seqn8}
\end{equation}

    The intrinsic stability condition are equivalent to requiring the convexity of the energy per particle in the single phase \cite{Ku04,Ku07,La07} by ignoring the finite size effects due to surface and Coulomb energies as shown in following. Here the $P=P^b+P^e$ is the total pressure of the npe system with the contributions $P^b$ and $P^e$ from baryons and electrons, respectively. The proton fraction $x_p=\frac{Q}{B}=\frac{\rho_p}{\rho}$ where $\rho={\rho_n}+{\rho_p}$ and the asymmetry parameter $X=\frac{\rho_n-\rho_p}{\rho_n+\rho_p}$. Total energy $\epsilon=\epsilon_b(x_p)+\epsilon_e(\mu)$.

\begin{eqnarray}
P=-\frac{\partial \epsilon}{\partial v}={\rho^2}\frac{\partial \epsilon}{\partial \rho}
\label{seqn9}
\end{eqnarray}

\begin{eqnarray}
(\frac{\partial P}{\partial v})_\mu=\frac{\partial P^b(\rho,x_p)}{\partial v}+\frac{\partial P^e(\mu)}{\partial v}
\label{seqn10}
\end{eqnarray}

Here $\frac{\partial P^e(\mu)}{\partial v}=0$ because if $\beta$-equilibrium is satisfied then $\mu= {\mu_n}-{\mu_p}=\mu_e$ and the electron contribution to $P^e$ is only a function of the chemical potential $\mu$ and in that case $(\frac{\partial P^e(\mu)}{\partial v})=0$. Eventually $-(\frac{\partial P}{\partial v})_\mu >0$ can be written as $-(\frac{\partial P^b}{\partial v})_\mu>0$.

\begin{eqnarray}
(\frac{\partial P}{\partial v})_\mu =\frac{\partial P^b}{\partial \rho}\frac{\partial \rho}{\partial v}+\frac{\partial P^b}{\partial x_p}\frac{\partial x_p}{\partial v}\nonumber\\
=-{\rho^2}\frac{\partial P^b}{\partial \rho}-{\rho^2}\frac{\partial P^b}{\partial x_p}\frac{\partial x_p}{\partial \rho}
\label{seqn11}
\end{eqnarray}

\begin{eqnarray}
-(\frac{\partial P}{\partial v})_\mu=\rho^2(\frac{\partial P^b}{\partial \rho}+\frac{\partial P^b}{\partial x_p}\,
\frac{\partial x_p}{\partial\rho})
\label{seqn12}
\end{eqnarray}

\begin{eqnarray}
\mu=\mu_n-\mu_p=-(\frac{\partial \epsilon^b}{\partial x^p})_\rho =-\frac{\partial \epsilon^b(\rho,x_p)}{\partial x_p}
\label{seqn13}
\end{eqnarray}
Differentiating above equation with respect to $x_p$ we get

\begin{equation}
\frac{\partial \mu}{\partial x_p}=-\frac{\partial^2 \epsilon^b}{\partial x^2_p}
\label{seqn14}
\end{equation}
From Eq.(9) we get

\begin{equation}
P^b={\rho^2}\frac{\partial \epsilon^b}{\partial \rho} 
\label{seqn15}
\end{equation}
and differentiating above with respect to $x_p$ one obtains

\begin{eqnarray}
(\frac{\partial P^b}{\partial x_p})=\rho^2{\frac{\partial^2 \epsilon^b}{\partial x_p \partial \rho}}={\rho^2} \epsilon^b_{\rho x_p}
\label{seqn16}
\end{eqnarray}
By Maxwell's relation

\begin{eqnarray}
(\frac{\partial x_p}{\partial \rho})_\mu=-{v^2}(\frac{\partial x_p}{\partial v})_\mu={v^2}(\frac{\partial P^b}{\partial \mu})_{s,v}
\label{seqn17}
\end{eqnarray}

\begin{eqnarray}
\frac{\partial P^b}{\partial \mu}=\frac{\frac{\partial P^b}{\partial x_p}}{\frac{\partial \mu}{\partial x_p}}=\frac{\rho^2\frac{\partial^2 \epsilon^b}{\partial \rho \partial x_p}}{\frac{\partial \mu}{\partial x_p}}= -\frac{\rho^2\frac{\partial^2 \epsilon^b}{\partial \rho \partial x_p}}{\frac{\partial^2 \epsilon^b}{\partial x_p^2}}
\label{seqn18}
\end{eqnarray}
Using Eq.(17) and Eq.(18) we get

\begin{eqnarray}
(\frac{\partial x_p}{\partial \rho})=-v^2 \rho^2\frac{\frac{\partial^2 \epsilon^b}{\partial \rho \partial x_p}}{\frac{\partial^2 \epsilon^b}{\partial x_p^2}} =-\frac{\frac{\partial^2 \epsilon^b}{\partial \rho \partial x_p}}{\frac{\partial^2 \epsilon^b}{\partial x_p^2}}
\label{seqn19}
\end{eqnarray}
From Eq.(15) 

\begin{eqnarray}
\frac{\partial P^b}{\partial \rho}=2\rho\frac{\partial \epsilon^b}{\partial \rho}+\rho^2\frac{\partial^2 \epsilon^b}{\partial \rho^2}
\label{seqn20}
\end{eqnarray}
Using Eq.(16), Eq.(19) and Eq.(20) in Eq.(12) we get

\begin{eqnarray}
-(\frac{\partial P^b}{\partial v})_\mu=\rho^2( 2\rho\frac{\partial \epsilon^b}{\partial \rho}+\rho^2\frac{\partial^2 \epsilon^b}{\partial \rho^2}-\rho^2\frac{\epsilon^b_{\rho x_p} \epsilon^b_{\rho x_p}}{\epsilon_{x_px_p}})
\label{seqn21}
\end{eqnarray}
The quantity $V_{thermal}$ which determines the thermodynamic instability region of neutron star matter at $\beta$-equilibrium is given by $V_{thermal}=-(\frac{\partial P}{\partial v})_\mu$. Hence

\begin{eqnarray}
V_{thermal}=\rho^2(2\rho\frac{\partial \epsilon^b}{\partial \rho}+\rho^2\frac{\partial^2 \epsilon^b}{\partial \rho^2}-\rho^2\frac{\epsilon^{b2}_{\rho x_p}}{\epsilon_{x_px_p}})
\label{seqn22}
\end{eqnarray}
The condition for core-crust transition is obtained by making $V_{thermal}=0$. In the following we drop the superscript $b$ and use $\epsilon$ for $\epsilon^b$ and $P$ for $P^b$.

\noindent
\section{Theoretical calculations}
\label{section3}

    The $\beta$-equilibrated nuclear matter EoS is obtained by evaluating the asymmetric nuclear matter EoS at the isospin asymmetry $X$ determined from the $\beta$-equilibrium proton fraction $x_p$ [$=\frac{\rho_p}{\rho}$], obtained approximately by solving 

\begin{equation}
 \hbar c (3 \pi^2\rho x_p)^{1/3}= 4E_{sym}(\rho) (1 - 2 x_p),
\label{seqn23}
\end{equation}
\noindent
where $E_{sym}(\rho)$ is the nuclear symmetry energy. In general $E_{sym}(\rho)$ is defined as $\frac{1}{2} \frac{\partial^2\epsilon(\rho,X)}{\partial{X^2}} \mid_{X=0}$. The higher-order terms in $X$ are negligible and to a good approximation, $E_{sym}(\rho)$=$\epsilon(\rho,1) -\epsilon(\rho,0)$ \cite{Kl06} which represents a penalty levied on the system as it departs from the symmetric limit of equal number of protons and neutrons and can be defined as the energy required per nucleon to change the SNM to pure neutron matter (PNM).

    The exact way of obtaining $\beta$-equilibrium proton fraction is by solving 

\begin{equation}
 \hbar c (3 \pi^2\rho x_p)^{1/3} = -\frac{\partial \epsilon(\rho,x_p)}{\partial x_p} = +2\frac{\partial \epsilon}{\partial X},
\label{seqn24}
\end{equation}
\noindent
where isospin asymmetry $X=1-2x_p$.

    The pressure $P$ of pure neutron matter (PNM) and $\beta$-equilibrated neutron star matter are plotted in Fig.-1 as functions of $\rho/\rho_0$. The continuous line represents the PNM and the dashed line (almost merges with the continuous line) represents the $\beta$-equilibrated neutron star matter (present calculations) whereas the dotted line represents the same using the A18 model using variational chain summation (VCS) of Akmal et al. \cite{Ak98} for the PNM. The areas enclosed by the continuous and the dashed lines in Fig.-1 correspond to the pressure regions for neutron matter consistent with the experimental flow data after inclusion of the pressures from asymmetry terms with weak (soft NM) and strong (stiff NM) density dependences, respectively \cite{Da02}. Although, the parameters of the density dependence of DDM3Y interaction have been tuned to reproduce $\rho_0$ and $\epsilon_0$ which are obtained from finite nuclei, the agreement of the present EoS with the experimental flow data, where the high density behaviour looks phenomenologically confirmed, justifies its extrapolation to high density. It is interesting to note that the RMF-NL3 incompressibility for SNM is 271.76 MeV \cite{La97,La99} which is about the same as 274.7$\pm$7.4 MeV obtained from the present calculation but the plot of $P$ versus $\rho/\rho_0$ for PNM of RMF using NL3 parameter set \cite{La97} does not pass through the pressure regions for neutron matter consistent with the experimental flow data \cite{Da02}.
    
    In Fig.-2 it can be seen that the maximum of the $\beta$-equilibrium proton fraction $x_p\sim0.0436$ calculated using the symmetry energy (approximate calculation) occurs at $\rho\sim1.35\rho_0$ whereas the exact calculation yields a maximum of $x_p\sim0.0422$ around the same density. Since the equilibrium proton fraction is always less than 1/9 \cite{La91} calculated value of $x_p$ forbids the direct URCA process. This feature is consistent with the fact that there are no strong indications \cite{AWS06,Ca06} that fast cooling occurs. It was also concluded theoretically that an acceptable EoS of asymmetric nuclear matter shall not allow the direct URCA process to occur in neutron stars with masses below 1.5 solar masses \cite{Kl06}. Even recent experimental observations that suggest high heat conductivity and enhanced core cooling process indicating the enhanced level of neutrino emission, were not attributed to the direct URCA process but were proposed to be due breaking and formation of neutron Cooper pairs \cite{Cr10,Da11,Dm11,Sh11}. 

    The intrinsic stability condition of a single phase for locally neutral matter under $\beta$-equilibrium is determined, thermodynamically, by the positivity of the $V_{thermal}$, under constant chemical potential which is generally valid in our case. However, the limiting density that breaks these conditions will correspond to the core-crust (liquid-solid) phase transition. Thus transition density $\rho_t$ (with corresponding pressure P$_t$ and proton fraction x$_{p(t)}$) is determined at which $V_{thermal}$ becomes zero and goes to negative with decreasing density. 
 
\begin{figure}[t]
\vspace{0.0cm}
\eject\centerline{\epsfig{file=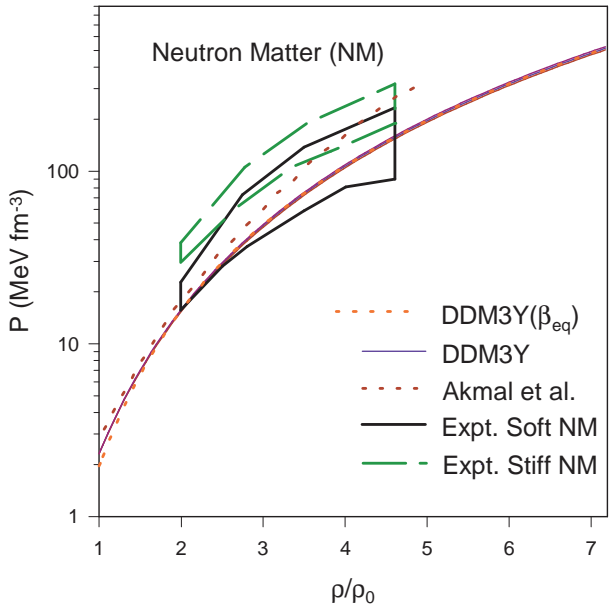,height=8cm,width=8cm}}
\caption
{Plots for pressure P of dense nuclear matter as functions of $\rho/\rho_0$. The continuous line represents the pure neutron matter and the dashed line represents the $\beta$-equilibrated neutron star matter. The dotted line represents the same for A18 model using variational chain summation (VCS) of Akmal et al. \cite{Ak98}. The areas enclosed by the continuous and the dashed lines correspond to the pressure regions for neutron matter consistent with the experimental flow data after inclusion of the pressures from asymmetry terms with weak (soft NM) and strong (stiff NM) density dependences, respectively \cite{Da02}.}
\label{fig1}
\vspace{0.0cm}
\end{figure}
\noindent 

\begin{figure}[t]
\vspace{0.0cm}
\eject\centerline{\epsfig{file=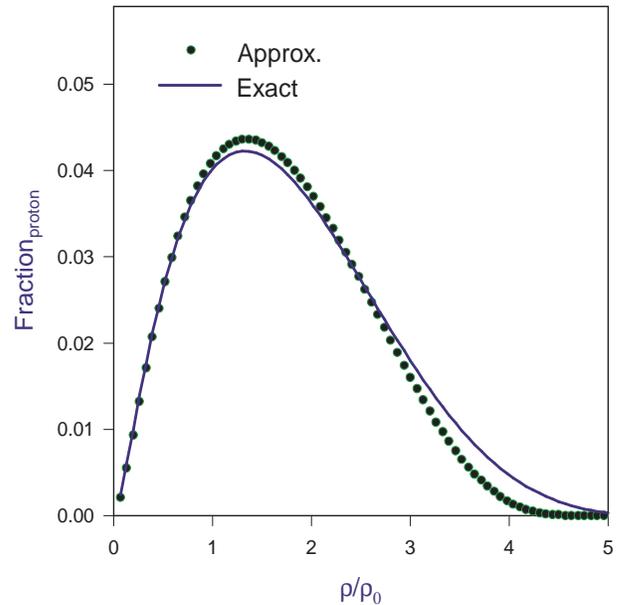,height=8cm,width=8cm}}
\caption
{The $\beta$ equilibrium proton fraction obtained from nuclear symmetry energy (approx.) and from exact calculations using DDM3Y interaction are plotted as functions of $\rho/\rho_0$.}
\label{fig2}
\vspace{0.0cm}
\end{figure}
\noindent 
    
\noindent
\section{ Results and discussion }
\label{section4}
    
    The stability of the $\beta$-equilibrated dense matter in neutron stars is investigated and the location of the inner edge of their crusts and core-crust transition density and pressure are determined using the DDM3Y effective nucleon-nucleon interaction. The results for the transition density, pressure and proton fraction at the inner edge separating the liquid core from the solid crust of neutron stars are calculated and presented in Table-1 for $n$=2/3. The symmetric nuclear matter incompressibility $K_\infty$, nuclear symmetry energy at saturation density $E_{sym}(\rho_0)$, the slope $L$ and  isospin dependent part $K_\tau$ of the isobaric incompressibility are also tabulated since these are all in excellent agreement with the constraints recently extracted from measured isotopic dependence of the giant monopole resonances in even-A Sn isotopes, from the neutron skin thickness of nuclei, and from analyses of experimental data on isospin diffusion and isotopic scaling in intermediate energy heavy-ion collisions.
    
\begin{table}[htbp]
\centering
\caption{Results of the present calculations (DDM3Y) of symmetric nuclear matter incompressibility $K_\infty$, nuclear symmetry energy at saturation density $E_{sym}(\rho_0)$, the slope $L$ and  isospin dependent part $K_\tau$ of the isobaric incompressibility (all in MeV) \cite{Ba09} are tabulated along with the saturation density and the density, pressure and proton fraction at the core-crust transition for $\beta$-equilibrated neutron star matter.}
\begin{tabular}{cccc}
\hline
\hline
$K_\infty$&$E_{sym}(\rho_0)$&$L$&$K_\tau$ \\ 
\hline
 $274.7\pm7.4$&$30.71\pm0.26$&$45.11\pm0.02$&$-408.97\pm3.01$ \\ 
\hline
$\rho_0$&$\rho_t$& P$_t$ & x$_{p(t)}$ \\
\hline
0.1533 fm$^{-3}$&0.0938 fm$^{-3}$&0.5006 MeV fm$^{-3}$& 0.0308 \\
\hline
\hline
\end{tabular} 
\label{table1}
\end{table}
\noindent 

\begin{table*}[htbp]
\centering
\caption{Variations of the core-crust transition density, pressure and proton fraction for $\beta$-equilibrated neutron star matter, symmetric nuclear matter incompressibility $K_\infty$ and isospin dependent part $K_\tau$ of isobaric incompressibility with parameter $n$.}
\begin{tabular}{||c|c|c|c|c|c||}
\hline
\hline
$n$&$\rho_t$& P$_t$ & x$_{p(t)}$&$K_\infty$&$K_\tau$ \\
\hline
&&&&& \\ 
1/6&0.0797 fm$^{-3}$&0.4134 MeV fm$^{-3}$& 0.0288&182.13 MeV&-293.42 MeV \\ \hline
&&&&&\\ 
1/3&0.0855 fm$^{-3}$&0.4520 MeV fm$^{-3}$& 0.0296&212.98 MeV&-332.16 MeV \\ \hline
&&&&&\\ 
1/2&0.0901 fm$^{-3}$&0.4801 MeV fm$^{-3}$& 0.0303&243.84 MeV&-370.65MeV \\ \hline
&&&&&\\ 
2/3&0.0938 fm$^{-3}$&0.5006 MeV fm$^{-3}$& 0.0308&274.69 MeV&-408.97 MeV \\ \hline
&&&&&\\ 
1  &0.0995 fm$^{-3}$&0.5264 MeV fm$^{-3}$& 0.0316&336.40 MeV&-485.28 MeV \\ \hline
\hline
\end{tabular}
\label{table2} 
\end{table*}
\noindent 

    It is recently conjectured that there may be a good correlation between the core-crust transition density and the symmetry energy slope $L$ and it is predicted that this behaviour should not depend on the relation between $L$ and $K_\tau$ \cite{Du10}. On the contrary, no correlation of the transition pressure with $L$ was obtained \cite{Du10}. In Table-2, variations of different quantities with parameter $n$ which controls the nuclear matter incompressibility are listed. It is worthwhile to mention here that the incompressibility increases with $n$. The standard value of $n$=2/3 used here has a unique importance because then the constant of density dependence $\beta$ has the dimension of cross section and can be interpreted as the isospin averaged effective nucleon-nucleon interaction cross section in ground state symmetric nuclear medium. For a nucleon in ground state nuclear matter $k_F\approx$ 1.3 fm$^{-1}$ and $q_0 \sim \hbar k_F c \approx$ 260 MeV and the present result for the `in medium' effective cross section is reasonably close to the value obtained from a rigorous Dirac-Brueckner-Hartree-Fock \cite{Sa06} calculations corresponding to such $k_F$ and $q_0$ values which is $\approx$ 12 mb. Using the value of $\beta$=1.5934 fm$^2$ along with the nucleonic density 0.1533 fm$^{-3}$, the value obtained for the nuclear mean free path $\lambda$ is about 4 fm which is in excellent agreement with that obtained using another method  \cite{Si83}. 

\noindent
\section{ Summary and conclusion }
\label{section5}

    In summary, the stability of the $\beta$-equilibrated dense nuclear matter is analyzed with respect to the thermodynamic stability conditions. The proton fraction obtained using nuclear symmetry energy does not affect seriously the results of exact calculation. Since the higher-order symmetry-energy coefficients are needed to describe reasonably well the proton fraction of the $\beta$-stable (npe) matter at high nuclear densities and the core-crust transition density \cite{Se14}, exact calculations are performed using the density dependent M3Y effective nucleon-nucleon interaction for investigating the proton fraction in neutron stars and the location of the inner edge of their crusts and their core-crust transition density and pressure. 
    
    The nucleon-nucleon effective interaction used in the present work, which is found to provide a unified description of elastic and inelastic scattering, various radioactivities and nuclear matter properties, also provides an excellent description of the $\beta$-equilibrated neutron star matter which is stiff enough at high densities to reconcile with the recent observations of the massive compact stars \cite{Ch10,Mi12,Ba14} while the corresponding symmetry energy is supersoft as preferred by the FOPI/GSI experimental data. The density, the pressure and the proton fraction at the inner edge separating the liquid core from the solid crust of the neutron stars determined to be $\rho_t=$ 0.0938 fm$^{-3}$, P$_t=$ 0.5006 MeV fm$^{-3}$ and x$_{p(t)}=$ 0.0308, respectively, are also in close agreement with other theoretical calculations \cite{Se14} corresponding to high nuclear incompressibility and with those obtained using SLy4 interaction \cite{Mo12}.

\end{document}